\documentclass[manuscript,screen]{acmart} %
\usepackage{graphicx}
\usepackage{makecell}
\usepackage{multirow}
\usepackage{soul}
\usepackage{tabularx}
\usepackage{titlesec}

\usepackage{comment}
\usepackage{ifthen}
\usepackage{xspace}
\usepackage{etoolbox}
\newtoggle{comments} 
\toggletrue{comments}

\usepackage{xcolor}
\usepackage[normalem]{ulem} 

\newif\ifshowrevisions
\showrevisionsfalse 

\ifshowrevisions
  \newcommand{\rev}[1]{\textcolor{blue}{#1}}
  \newcommand{\revdel}[1]{\textcolor{red}{\sout{#1}}}
\else
  \newcommand{\rev}[1]{#1}
  \newcommand{\revdel}[1]{}
\fi

\AtBeginDocument{%
  \providecommand\BibTeX{{%
    \normalfont B\kern-0.5em{\scshape i\kern-0.25em b}\kern-0.8em\TeX}}}

\usepackage{array}
\usepackage{booktabs}
\usepackage{geometry}
\usepackage{adjustbox}
\usepackage{longtable}
\usepackage[normalem]{ulem}

\setcopyright{acmcopyright}
\copyrightyear{2018}
\acmYear{2018}
\acmDOI{XXXXXXX.XXXXXXX}

\acmConference[CHI'26]{CHI'2026}
\acmPrice{15.00}
\acmISBN{978-1-4503-XXXX-X/18/06}

\newcommand{\elide}[1]{\textelp{}} 

\begin{document}

\title{“My Brother is a School Principal, Earns About \$80,000 Per Year... But When the Kids See Me, ‘Wow, Uncle, You have 1500 Followers on TikTok!’”: A Study of Blind TikTokers’ Alternative Professional Development Experiences}

\author{Yao Lyu}
\orcid{0000-0003-3962-4868}
\affiliation{%
  \institution{University of Michigan}
  \city{Ann Arbor}
  \state{Michigan}
  \country{USA}
}
\email{yaolyu@umich.edu}

\author{Tawanna Dillahunt}
\orcid{0000-0002-1155-8507}
\affiliation{%
  \institution{University of Michigan}
  \city{Ann Arbor}
  \state{Michigan}
  \country{USA}
}
\email{tdillahu@umich.edu}

\author{Jiaying Liu}
\orcid{0000-0002-5398-1485}
\affiliation{%
  \institution{University of Texas}
    \city{Austin}
  \state{Texas}
  \country{USA}
}
\email{jiayingliu@utexas.edu}

\author{John M. Carroll}
\orcid{0000-0001-5189-337X}
\affiliation{%
  \institution{Pennsylvania State University}
  \city{University Park}
  \state{Pennsylvania}
  \country{USA}
}
\email{jmcarroll@psu.edu}

\renewcommand{\shortauthors}{}


\begin{abstract}

\rev{One’s profession is an essential part of modern life. Traditionally, professional development has been criticized for excluding people with disabilities. People with visual impairments, for example, face disproportionately low employment rates, highlighting persistent gaps in professional opportunities. Recently, there has been growing research on social media platforms as spaces for more equitable career development approaches. In this paper, we present an interview study on the professional development experiences of 60 people with visual impairments on TikTok (also known as “BlindTokers”). We report BlindTokers’ goals, strategies, and challenges, supported by detailed examples and in-depth analysis. Based on the findings, we identified that BlindTokers’ practices reveal an \textit{\textbf{alternative professional development}} approach that is more flexible, inclusive, personalized, and diversified than traditional models. Our study also extends professional development research by foregrounding emerging digital skills and proposing design implications to foster more equitable and inclusive professional opportunities.}

\end{abstract}




\maketitle

\section{Introduction}
One's profession is an essential part of modern life, closely tied to social recognition, career growth, and economic well-being. \rev{Professional development, which refers to the self-directed process of pursuing professional growth, including but not limited to skill enhancement, visibility increase, network expansion, revenue generation, or career advancement \cite{guskey_professional_2002,dachner_future_2021}, is of great importance.} However, the power to define profession, including training, skills, and qualifications has long been concentrated in institutional hands \cite{greenwood_attributes_1957, cogan_problem_1955, freidson_professionalism_1994, freidson_professionalism_2001}. This institutionalized conception of professional development is problematic, as the monopolization of power contributes to the marginalization of people with disabilities, such as limited opportunities \cite{schur_corporate_2005}, stigmatization \cite{goffman_stigma_2009}, and inaccessibility \cite{anand_role_2017}. \rev{Take people with disabilities as an example: The U.S. Bureau of Labor Statistics reported that only 22.7\% of people with disabilities were employed in 2024 \cite{bureau_of_labor_statistics_persons_2025}. According to the 2023 U.S. Census Bureau, 52.3\% of working-age people with visual impairments were employed, compared to 76.3\% of working-age people in the general population \cite{us_census_bureau_employment_2023}. This disparity in employment rates underscores the unequal access to professional development faced by people with disabilities \cite{bonaccio_participation_2020}.}  



Human-Computer Interaction (HCI) and accessibility researchers have conducted extensive work \cite{mountapmbeme_addressing_2022, thoo_large-scale_2023} on revealing people with disabilities' unequal access to professional opportunities \cite{chen_silent_2024, wang_invisible_2022, sannon_toward_2022} and proposing designs to address the inequality in professional contexts \cite{cha_you_2024,saha_tutoria11y_2023, heinz_dynamic_2021}. The contexts are various, including job seeking \cite{kong_understanding_2024}, productivity \cite{saha_understanding_2020, philips_helping_2024,kumar_uncovering_2024}, skill development \cite{gunupudi_scaffolding_2024}, education \cite{song_emobridge_2024, coverdale_digital_2024, lu_playing_2023,lu_why_2023}, professional learning \cite{alonzo_reading_2020, storer_its_2021}, collaboration \cite{potluri_codewalk_2022,pandey_understanding_2021}, and peer support \cite{saha_understanding_2023}. That said, professional development in the aforementioned studies focuses on traditional contexts, like companies and universities.

Recently, there has been a growing line of research on the impact of emerging sociotechnical contexts, such as social media platforms, on career development. Researchers have reported that social media platforms offer people greater flexibility in envisioning, planning, strategizing, and achieving professional growth \cite{ding_as_2022}. This flexibility, shaped by platforms’ socio-technical dynamics, allows users to explore varied career paths, receive evaluations through unconventional means, and earn income from multiple sources. In addition, social media can also provide job seekers with exposure to diverse career possibilities, inspirational content \cite{nguyen_rol,corvite_social_2024}. These studies demonstrate social media platforms' important role in transforming what professional development looks like in traditional contexts. 



\begin{figure}[htp]
    \centering
    \includegraphics[scale=0.43]{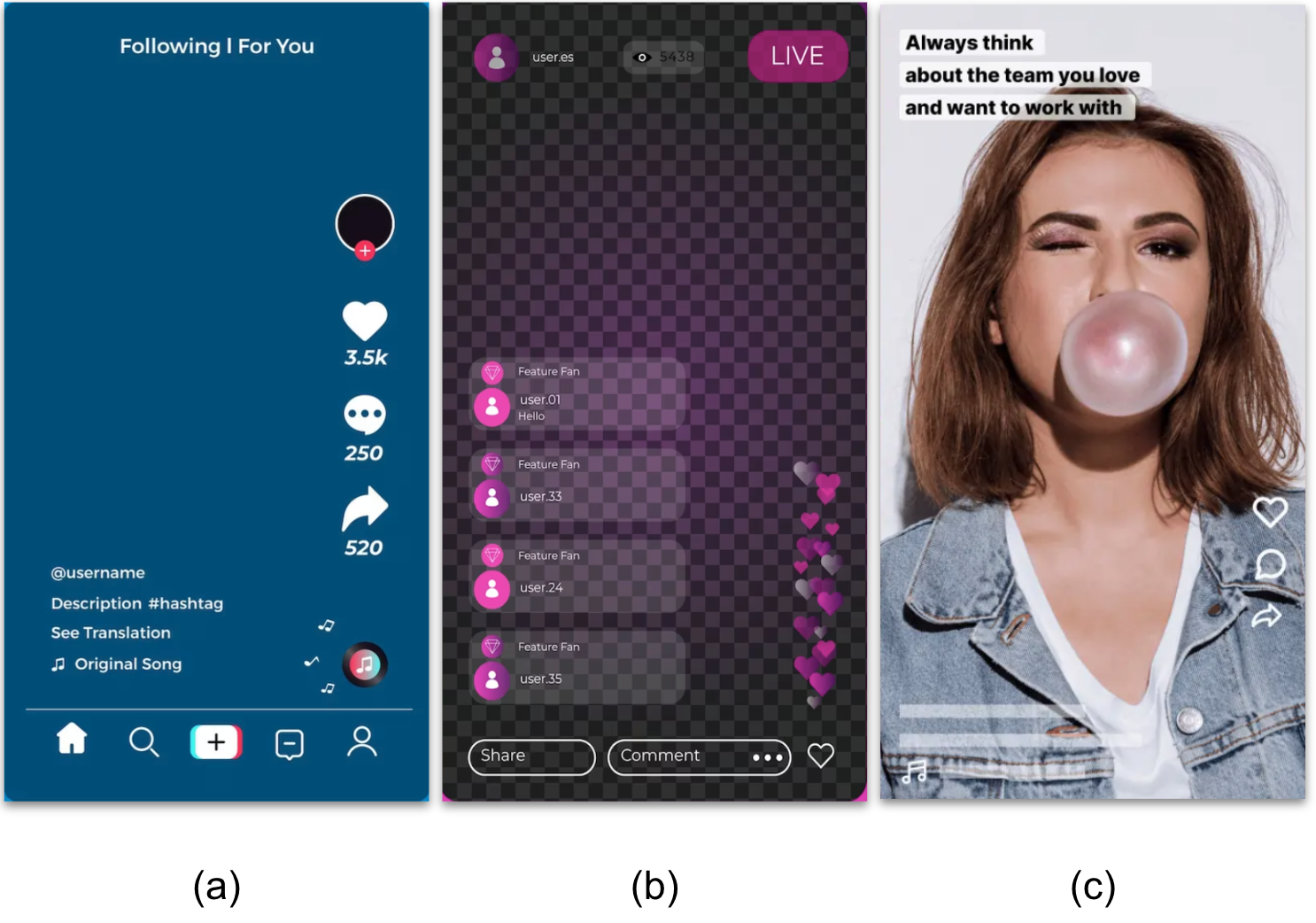}
    \Description{TikTok’s three interfaces are displayed in a combined image. (a) The left panel shows the main feed with video engagement metrics (likes, comments, shares), user information, hashtags, and music details. (b) The middle panel depicts a livestream screen with user comments, viewer count, and floating heart reactions. (c) The right panel features a video post of a woman blowing bubble gum, overlaid with motivational text and engagement icons.}
    \caption{Three key TikTok interfaces: (a) The main feed displays short videos with interactive buttons (like, comment, share), user info, hashtags, and music credits. (b) The livestream screen shows real-time viewer engagement through comments, likes, and reactions. (c) A typical video post with text overlay, user interaction icons, and background music indicators.}
    \label{fig:tiktok}
\end{figure}

However, social media platforms' specific impact on people with disabilities' professional development is questionable. Numerous studies on people with disabilities on social media platforms have revealed the issues of accessibility and inclusiveness that significantly hinder users' experiences \cite{lyu_i_2024,heung_vulnerable_2024,choi__2022, low_twitter_2019, seo_exploring_2017}. These barriers may further impose additional burdens on people with disabilities \cite{rong_it_2022} and hence affect their professional development. Therefore, a comprehensive and in-depth study on people with disabilities' experience of professional development on social media platforms is warranted. In this paper, we introduce an interview study on the professional development experience of people with visual impairments on TikTok (also known as "BlindTokers"). TikTok is a visually driven platform characterized by short-form video content, hashtag-based community structures, and algorithmically curated content feeds (shown in Figure-\ref{fig:tiktok}). This study is part of a long-term ethnography project investigating the blind community on TikTok. Professional development emerged as a key theme midway through the project and inspired the current study (see the Methods section for more details). Specifically, we ask three research questions: 


\begin{itemize}
\item RQ1: What are the primary professional goals for BlindTokers?
\item RQ2: What strategies do BlindTokers use to leverage for the goals?
\item RQ3: What challenges do BlindTokers encounter on TikTok?
\end{itemize}

We interviewed 60 BlindTokers and used thematic analysis to understand their perceptions and practices related to professional development. The results uncover how BlindTokers pursue both traditional and nontraditional professional paths, expanding professional development beyond institutional settings (e.g., professional paths span from church minister to professional content creators). Throughout the study, we observed that BlindTokers’ professional development activities involved efforts to create career opportunities through learning, training, networking, and, in some cases, monetization. Specifically, our findings addressed three research questions through three key themes. \rev{First, we described BlindTokers’ goals for professional development, including trying to avoid unemployment and improve their career status. These goals often happen at the same time, as many participants expressed their desire for getting or keeping a job while also hoping to move forward in their careers with the help of TikTok.} Second, we found that their strategies for professional development on TikTok included monetizing for profit, honing skills for competency, and socializing to build networks. Third, we identified two main types of challenges, technical and social, that significantly hindered BlindTokers’ professionalization experiences.

Based on the findings, we first explored BlindTokers' professional development on social media platforms, unpacking participants' understandings and strategies for leveraging TikTok’s socio-technical environment to pursue more diversified forms of professional growth. We identified an \textbf{alternative professional development} approach, which helps BlindTokers gain more control, reframes visual impairments as a competency, is characterized by individualized interpretations, and enables diversified revenue streams. Second, we highlight the professional skill sets that BlindTokers exhibited within this redefined professional development approach, including accessibility work, algorithmic literacy, advocacy skills, and emotional labor. Lastly, we provide specific design implications to support BlindTokers' alternative approach to professional development. These include the recognition of BlindTokers' accessibility work by professional stakeholders, the adoption of more equitable monetization models by sponsors, the construction of safe and inclusive networking spaces by the TikTok platform, and the sharing of resources by public sectors on TikTok. The contributions of this study are multifaceted: (1) We contribute empirical findings of BlindToker's use of the platform for professional development. Our empirical results reveal the diverse understanding and practices of people with visual impairments regarding professional development, which is shaped by platform designs and policies. (2) We extend professional development literature by introducing an alternative professional development approach, which is more inclusive, flexible, and personalized. (3) We foreground emerging professional skills in this new context and propose design implications to support this alternative approach.

\section{Related Work}

\subsection{Professional Development and People with Visual Impairments}

\rev{Profession is "a vocation in which professed knowledge of some branch of learning is used in its application to the affairs of others, or in the practice of an art based upon it [p. 656]" \cite{hughes_professions_1963}. Profession plays an important role in modern society as it "delivers esoteric services—advice or action or both—to individuals, organizations or government...Even when manual, the action...is determined by esoteric knowledge systematically formulated and applied to problems of a client. [p. 655]" \cite{hughes_professions_1963}. Professional development refers to the self-directed process of pursuing professional growth, including but not limited to skill enhancement, visibility increase, network expansion, revenue generation, or career advancement \cite{guskey_professional_2002,dachner_future_2021}.} 

\rev{Traditionally, the definition of professional has been institutional. As emphasized by Ernest Greenwood \cite{greenwood_attributes_1957}, an American Sociologist, the profession convinces society to "institute in its behalf a licensing system for screening those qualified to practice professional skill \cite{greenwood_attributes_1957} [p. 49]." He also revealed that the powers and privileges of licensing professionals "constitute a monopoly... to the professional group \cite{greenwood_attributes_1957} [p. 49]." Under this institutional model, professional development is understood as the process of meeting the training requirements, skill standards, and credentialing criteria established by these gatekeeping bodies. Therefore, professional development, where people seek qualification and advancement in professions, is also defined in institutional terms.}

The institutionalized conception of professional development is problematic because the concentration of power within this sphere marginalizes people with disabilities. Extensive work has revealed that in the current system,  people with disabilities are experiencing limited professional opportunities \cite{qiu_impact_2023,fisher_job_2016}. Accessibility researchers have actively challenged ableist norms in professional development. This body of work emphasizes the importance of applying equitable models when assessing the professional performance of individuals with disabilities \cite{samuels_six_2017, kafer_feminist_2013}. For example, Alison Kafer \cite{kafer_feminist_2013} introduced the concept of "crip time" to resist the normative, ableist view of time, which is often rigid, linear, and productivity-driven. Instead, "Crip Time" advocates for a more inclusive understanding of time that reflects the needs, rhythms, and realities of people with disabilities. 

HCI researchers have been addressing the challenges faced by people with disabilities in professional development through socio-technical approaches. Their extensive research has uncovered the unequal access to professional opportunities experienced by people with disabilities \cite{mountapmbeme_addressing_2022, thoo_large-scale_2023} and proposed designs to address this inequality across various professional contexts \cite{chen_silent_2024, wang_invisible_2022, sannon_toward_2022}. These contexts include job seeking \cite{kong_understanding_2024}, productivity \cite{saha_understanding_2020, philips_helping_2024, kumar_uncovering_2024}, skill development \cite{gunupudi_scaffolding_2024}, education \cite{song_emobridge_2024, coverdale_digital_2024, lu_playing_2023, lu_why_2023}, professional learning \cite{alonzo_reading_2020, storer_its_2021}, collaboration \cite{potluri_codewalk_2022, pandey_understanding_2021}, and peer support \cite{saha_understanding_2023}. In addition to revealing unequal access to professional opportunities, researchers have highlighted the perpetuation of ableist standards within these contexts \cite{mack_anticipate_2022, borsotti_neurodiversity_2024, saha_understanding_2020}. For instance, Saha and Piper's study on professionals with visual impairments in audio production settings \cite{saha_understanding_2020} argued that employers, by failing to provide accessible work environments and applying one-size-fits-all productivity standards, were reinforcing ableist norms. The study also called for future research to reconsider the standards applied in professional contexts.

That said, while these studies focus on professional development in traditional contexts such as companies and universities, limited work has explored professional development in emerging contexts, such as platforms. These platforms have been shown to offer greater flexibility in addressing situated professional needs \cite{ding_as_2022}. To fill this gap, we conducted a study on the professional development of people with visual impairments on one of the most popular platforms, TikTok. Through this study, we investigate BlindTokers' perceptions and practices toward professional growth, shaped by TikTok's socio-technical designs. Our goal is to explore and identify alternative approaches to professional development and contribute to the literature by offering a more inclusive definition of professional development.

\subsection{Professional Development and Social Media Platforms}

Social media platforms have become an increasingly important component of professional development. The vast amount of information and opportunities for information exchange on platforms like YouTube, TikTok, and Instagram allow users to learn, practice professional skills, and network, key elements of professional growth \cite{freidson_professionalism_2001}. While these platforms can serve as valuable external resources for professional development, they are also playing an increasingly central role in many careers, particularly for content creators, whose primary professional work revolves around generating content for their audience \cite{simpson_rethinking_2023, ma_multi-platform_2023,borgos-rodriguez_understanding_2023,rechkemmer_understanding_2022}. For example, Corvite and Hui \cite{corvite_social_2024} presented an interview study on 19 job seekers who utilized various social media platforms like Instagram, Facebook, YouTube, and TikTok, for career envisioning. The study found that social media posts helped the interviewees increase their awareness of possible professional pathways and motivate them to take action for desired careers. In addition, Ding et al. \cite{ding_as_2022} interviewed 13 Bilibili.com users on their transition from casual users to professional content creators. This process involves navigating and aligning with the socio-technical environment of the Bilibili platform, including understanding and acquiring technical skills and developing and sharing the moral requirements as professional uploaders. They used "individualized professionalization" to refer to the individualized perspective, underscoring how digital platforms enable creators to professionalize on their own terms while fostering shared standards within their communities.

A key theme in research on professional development on social media platforms is the recognition of professional content creators’ diverse forms of labor, work that is often unrecognized, invisible, or ignored \cite{branham_invisible_2015, simpson_hey_2023, chen_voice_2025,das_that_2024,akter_if_2023}. Acknowledging these types of labor not only deepens our understanding of professionals' experiences but also identifies important criteria for more comprehensive evaluations of professional performance. Simpson and Semaan \cite{simpson_rethinking_2023} studied the content creation routines of 15 professional TikTokers, using the term creative labor to describe “the work involved to professionalize, monetize, make visible, and relate to one's audience as a cultural producer” \cite{simpson_rethinking_2023} [p. 3]. This conceptualization emphasizes creators’ ongoing engagement with their audience communities, who serve as vital sources of learning, networking, and monetization. Ma and Kou \cite{ma_how_2021} examined posts in a Reddit community where YouTubers discussed their experiences navigating YouTube’s algorithms. They introduced the notion of algorithmic labor, which "involves striking a delicate balance between enhancing visibility on the one hand and avoiding moderation algorithms on the other" \cite{ma_how_2021} [p. 18]. Additionally, few scholars have explored the labor of content creators from marginalized communities. For instance, Lyu et al. \cite{lyu_i_2024} studied 45 visually impaired participants on short video platforms. Their research focused on collective identity work and highlighted its positive aspects. They coined the term flourishing labor to describe "efforts made by [participants] that contribute to the flourishing [of the community]... particularly in helping them realize their desires for pleasure, personal significance, and virtue" \cite{lyu_i_2024} [p. 17].

However, there remains limited research on the professional development of people with visual impairments on social media platforms, particularly regarding their motivations, strategies, and challenges throughout this process. To address this gap, we explore how BlindTokers engage in professional development work. Specifically, we examine the various forms of labor related to professional training, learning, networking, and monetization that BlindTokers undertake. Through our findings, we aim to contribute to the literature by offering a feasible framework for recognizing and acknowledging the professional work of BlindTokers.

\subsection{Social Media and People with Visual Impairments}
Our study is also relevant to the experiences of people with visual impairments using social media. On social media platforms, people with visual impairments enjoy others' stories, learn useful instructions for overcoming everyday accessibility issues, share their voices, create educational videos, and build communities \cite{lyu_because_2024,rong_it_2022,lyu_i_2024,lyu_i_2024-1}. The current literature on social media use by people with visual impairments focuses on three major themes: content consumption, content creation, and community construction \cite{lyu_systematic_2025,seo_challenges_2021,seo_understanding_2021, tigwell_emoji_2020,li_it_2022}.

Prior work on content consumption investigates how people with visual impairments browse, share, and interact with social media content. When consuming content, they are particularly concerned about accessibility across various modalities, including text \cite{low_twitter_2019}, images \cite{dos_santos_marques_audio_2017}, emojis \cite{tigwell_emoji_2020}, GIFs \cite{gleason_making_2020}, and videos \cite{seo_understanding_2021}. Researchers have also explored their nuanced requirements for accessing information in different contexts. Stangl et al. \cite{stangl_person_2020} examined how people with visual impairments interpret images depending on the platform: on general social networks, they cared about who was in the picture and what they were doing; on dating platforms, they focused more on appearance-related attributes such as hairstyle, body figure, and attractiveness.

Research on content creation has paid attention to how people with visual impairments develop and implement creative ideas, as well as the accessibility barriers that hinder these processes. They engage in a wide range of content creation activities on social media, such as composing music, creating videos, and hosting livestreams \cite{lyu_because_2024,lyu_i_2024}. In these scenarios, their use of technology goes beyond consuming information, it must also support their nuanced and creative practices. Lyu et al. \cite{lyu_i_2024} reported how users with visual impairments created content that reflected their blind identity, using specific filters and sound effects.

Network building on social media intersects with the prior two themes and focuses on how people with visual impairments interact to foster networks, often centered around shared experiences \cite{seo_exploring_2017, rong_it_2022}. These networks are built for several purposes, the most important being the creation of a community where individuals can connect and support one another. This support may include informational, financial, or emotional aid \cite{saha_understanding_2023,lyu_because_2024}. Another increasingly important reason for building such communities is to foster solidarity against ableism, systemic discrimination that manifests through exclusionary platform policies or harassment from sighted users \cite{lyu_i_2024, heung_vulnerable_2024}.

Across these three themes, two critical challenges consistently shape the experiences of people with visual impairments: accessibility and ableism. While accessibility issues arise from technical limitations of social media platforms \cite{zhang_ga11y_2022,stangl_person_2020}, ableism reflects broader social justice problems that exclude people with visual impairments \cite{saha_understanding_2023,lyu_because_2024}. These two major problems affect users with visual impairments' experience across various scenarios. However, little research has been done in the professional development context. To address this gap, we propose this study of BlindTokers' professional development. In our study, we join the research of people with visual impairments' use of social media, focusing on the new context of professional development. Through the study of BlindTokers, we want to examine how TikTok's socio-technical designs affect BlindTokers' specific practices for professional growth. We also aim to explore new challenges and design opportunities for TikTok or similar social media platforms in the new context.

\section{Methods}
\subsection{Study Background}
This study is part of a long-term ethnography project investigating the blind community on TikTok, which was approved by our Institutional Review Board (IRB). The overarching goal of the larger project was to understand how TikTokers with visual impairments use the platform, especially from an accessibility perspective. We adopted a reflexive approach throughout both data collection and analysis. Specifically, we began with a preliminary interview protocol but continuously revised it in response to participants' feedback and the stories they shared. For example, while our initial focus was on general platform use and accessibility issues, we noticed that participants frequently discussed using TikTok for networking, learning, and training. In response, we added a dedicated section to the interview protocol to explore these areas in greater depth.

For data collection, we conducted semi-structured interviews, covering not only general questions about participants' overall perceptions of TikTok but also specific topics such as how they use TikTok for professional learning, training, and networking. For the data analysis for this specific study, we employed thematic analysis methods to better understand participants' experiences, including their motivations for professional development, strategies for using TikTok to achieve it, and perceived challenges while pursuing professional growth. 

\subsection{Recruitment}

\rev{To recruit participants, we created official TikTok accounts, used our real identities, and clearly stated the purpose of the project. We searched for users who self-identified as blind or visually impaired by using keywords such as “blind”, “low vision”, “vision loss”, “visually impaired”, and “visual impairment” in their usernames, bios, or content descriptions. To reach out to participants and foster collaborative relationships with the community, we aimed to ensure transparency when introducing the main researchers and the project. We followed these users and actively engaged with their content, such as leaving supportive comments on their videos and attending their live streams. We also posted videos introducing our research team, our ongoing project, and aspects of our daily lives as researchers working with blind communities. Afterwards, some BlindTokers followed us back, enabling us to send private messages with more detailed project information. (In 2022-2023, TikTok only allowed private messages between users who mutually followed each other.) Before formally inviting participants to our study, we used private messages to introduce our team and project and answer any questions BlindTokers raised.}

\rev{We took this approach to build trust with potential participants and to demonstrate that we were a legitimate research team. Here, trust means not only confirming our authenticity as researchers affiliated with a real institution, but also showing that we were genuinely committed to understanding and supporting this community. Establishing mutual interest also helped increase participants’ responsiveness and encouraged some to assist with snowball sampling. This approach is informed by prior work on recruiting participants from online marginalized communities, such as rapport-building before recruitment \cite{haimson_social_2018}.}

Except for the interactions on TikTok, all other communication channels were based on participants’ preferences, such as email, phone calls, or Zoom, to ensure accessibility. To further reduce accessibility barriers, all data collection, including demographic information, was conducted through verbal communication. At the end of the interviews, we asked participants to recommend other TikTokers with visual impairments. This snowball sampling approach enabled us to reach out to approximately 500 blind TikTokers and successfully recruit 60 diverse participants between Summer 2022 and Spring 2023. 

\rev{We summarize participants' demographic information as follows. In terms of age, 16 participants were aged 18–25, 26 were aged 26–35, 10 were aged 36–45, 6 were aged 46-55, and 2 were aged 56–65. Speaking of gender, 34 participants self-identified as women, 22 self-identified as men, and 4 self-identified as non-binary people. Talking about visual impairment, 48 participants were legally blind, 8 had low vision, and 4 were totally blind. In terms of education, 17 participants completed high school or earned a General Education Degree, 17 had an associate’s degree, 14 held a bachelor’s degree, 10 had a master’s degree, 1 had a PhD, and 1 did not report their educational background. Fifty-two of the participants were based in the USA, four came from the UK, one from Bolivia, one from Canada, and two did not report their residency information. More detailed information about participants can be found in Table-\ref{table:demo}, and to protect their privacy, we do not list residency information in the table.}

\rev{We also collected participants' occupation and income information. Speaking of occupation, out of 60 participants, 10 were students, 13 participants were unemployed, 4 of them were professional TikTokers, and 33 were employed or had their own business. Their job experiences were diverse, covering areas like creativity, education, and technology. In terms of income, 23 participants who were students or unemployed had no income, and 10 did not report income information. Among the remaining 27 people, who were employed and reported their annual income, 21 earned between 10K and 100K per year, 2 earned between 100K and 500K per year, and 4 earned more than 500K per year (all in USD).} 

\rev{To be noted, there were among the participants who were employed, reported their income information, and lived in the U.S. (N=24), 17 of them had incomes lower than the average income per capita in the U.S. in 2023 (\$ 68,531/year) \cite{noauthor_per_2025}. These 17 participants had a large variety of jobs, like motivational speaker, church minister, sales manager, and tech expert. The other 7 who reported that their annual incomes were higher than the average income per capita in the U.S. had jobs like financial consultant, lawyer, car dealer, or high school teacher. Based on the current data, though limited, we can observe that participants' income status was not closely associated with their job types. For instance, many participants were in similar fields (e.g., P14 and P19 were in sales, P53 and P44 were teachers), but their incomes were drastically different. Also, owning a business is not a strong indicator of financial income. P13 and P11 both own their business, but their annual income was also very different (1,000K vs 40K).}

\begin{table}[]
\caption{Demographic Information of Participants}
\small
\label{table:demo}
\begin{tabular}{llllllll}
\hline 
No. & Age & Gender & Impairment    & Education & Occupation                & Income/Year     & Followers          \\
 &  &            &     &     &                  & in USD  &               \\
\hline
P01 & 26-35 & Male           & Low Vision    & Bachelor    & Student                 & N/A  & 1-10K              \\
P02 & 18-25 & Female         & Low Vision    & Associate   & Student                 & N/A  & 1-10K              \\
P03 & 18-25 & Female         & Legally Blind & High School & Student                 & N/A  & 1-10K              \\
P04 & 18-25 & Female         & Legally Blind & High School & Unemployed              & 0    & 1-10K              \\
P05 & 46-55 & Male           & Legally Blind & Bachelor    & Software Developer      & 150k & 10-100K            \\
P06 & 36-45 & Female         & Legally Blind & Associate   & Tech Advisor            & 37K  & 1-10K              \\
P07 & 26-35 & Female         & Legally Blind & Associate   & Unemployed              & 0    & 1-10K              \\
P08 & 26-35 & Male           & Legally Blind & High School & Musician                & N/A  & 10-100K            \\
P09 & 36-45 & Male           & Legally Blind & Master      & XR Designer             & 900K & 1-10K              \\
P10 & 26-35 & Female         & Low Vision    & Bachelor    & Musician                & 45K  & \textless{}1K      \\
P11 & 46-55 & Female         & Legally Blind & Bachelor    & Gallary Owner           & 40K  & \textless{}1K      \\
P12 & 26-35 & Male           & Legally Blind & High School & Unemployed              & 0    & 1-10K              \\
P13 & 56-65 & Female         & Totally Blind & High School & Finacial Consultant     & 1,000K   & \textless{}1K      \\
P14 & 36-45 & Male           & Low Vision    & High School & Car Dealer              & 500K & \textless{}1K      \\
P15 & 46-55 & Male           & Legally Blind & PhD         & Lawyer                  & 500K & \textless{}1K      \\
P16 & 36-45 & Male           & Legally Blind & Master      & Church Minister         & 20K  & \textless{}1K      \\
P17 & 26-35 & Male           & Legally Blind & High School & Musician                & N/A  & \textless{}1K      \\
P18 & 36-45 & Male           & Legally Blind & Associate   & Fitness Instructor      & 28K  & \textless{}1K      \\
P19 & 26-35 & Male           & Low Vision    & High School & Sales Manager           & 50K  & \textless{}1K      \\
P20 & 26-35 & Female         & Low Vision    & Associate   & Unemployed              & 0    & \textless{}1K      \\
P21 & 26-35 & Enby           & Legally Blind & High School & Unemployed              & 0    & \textless{}1K      \\
P22 & 26-35 & Female         & Legally Blind & Associate   & Student                 & N/A  & 10-100K            \\
P23 & 26-35 & Female         & Legally Blind & GED         & Therapist               & 135K & 1-10K              \\
P24 & 26-35 & Male           & Legally Blind & Bachelor    & Motivational Speaker    & 20K  & \textgreater{}100K \\
P25 & 26-35 & Male           & Legally Blind & Associate   & Tech Expert             & 60K  & 10-100K            \\
P26 & 18-25 & Male           & Legally Blind & Bachelor    & Digital Marketing       & 35K  & 1-10K              \\
P27 & 26-35 & Female         & Low Vision    & High School & Unemployed              & 0    & 1-10K              \\
P28 & 26-35 & Male           & Legally Blind & High School & Student                 & N/A  & \textless{}1K      \\
P29 & 46-55 & Female         & Legally Blind & GED         & Tech Advisor            & 22K  & 1-10K              \\
P30 & 18-25 & Enby           & Legally Blind & High School & Student                 & N/A  & 10-100K            \\
P31 & 26-35 & Female         & Legally Blind & Bachelor    & Unemployed              & 0    & \textless{}1K      \\
P32 & 18-25 & Male           & Legally Blind & Bachelor    & Unemployed              & 0    & \textless{}1K      \\
P33 & 26-35 & Female         & Legally Blind & Bachelor    & Unemployed              & 0    & 1-10K              \\
P34 & 18-25 & Enby           & Totally Blind & Bachelor    & Musician                & N/A  & 10-100K            \\
P35 & 26-35 & Male           & Totally Blind & N/A         & Civil Service           & 38K  & 10-100K            \\
P36 & 18-25 & Female         & Legally Blind & High School & Content Creator         & N/A  & \textless{}1K      \\
P37 & 18-25 & Female         & Legally Blind & Associate   & Unemployed              & 0    & \textgreater{}100K \\
P38 & 36-45 & Male           & Legally Blind & Associate   & Media                   & N/A  & \textless{}1K      \\
P39 & 26-35 & Female         & Legally Blind & Associate     & Data Entry              & N/A  & \textless{}1K      \\
P40 & 18-25 & Female         & Low vision    & Associate   & Graphic Designer        & N/A  & 1-10K              \\
P41 & 56-66 & Female         & Legally Blind & Associate   & Realtor                 & 50K  & \textless{}1K      \\
P42 & 26-35 & Female         & Legally Blind & GED         & Unemployed              & 0    & 1-10K              \\
P43 & 36-45 & Female         & Legally Blind & Masters     & Mental Health Counselor & 30K  & \textless{}1K      \\
P44 & 26-35 & Female         & Legally Blind & Masters     & Teacher                 & 40K  & 10-100K            \\
P45 & 26-35 & Female         & Legally Blind & Associate   & Accessibility Tester    & 35K  & 1-10K              \\
P46 & 18-25 & Female         & Legally Blind & Associate   & Unemployed              & 0    & 1-10K              \\
P47 & 26-35 & Female & Legally Blind & Associate   & Content Creator              & N/A  & 10-100K            \\
P48 & 18-25 & Male   & Legally Blind & High School & Student                 & N/A  & \textless{}1K      \\
P49 & 46-55 & Female         & Legally Blind & Master      & Store Owner             & 20k  & 10-100K            \\
P50 & 36-45 & Female         & Legally Blind & Master      & Unemployed              & 0    & 1-10K              \\
P51 & 36-45 & Female         & Legally Blind & Associate   & Federal Employee        & 60K  & 1-10K              \\
P52 & 36-45 & Female         & Legally Blind & Associate   & Content Creator              & 20K  & 1-10K              \\
P53 & 26-35 & Male           & Legally Blind & Master      & Teacher                 & 92K  & 10-100K            \\
P54 & 46-55 & Enby           & Legally Blind & Bachelor    & Actress                 & 25K  & \textless{}1K      \\
P55 & 26-35 & Female         & Legally Blind & Bachelor    & Self-employed Writer    & N/A  & N/A                \\
P56 & 18-25 & Female         & Legally Blind & Bachelor    & Content Creator       & N/A  & N/A                \\
P57 & 18-25 & Male   & Totally Blind & Bachelor    & Student                 & N/A  & 1-10K              \\
P58 & 26-35 & Male   & Legally Blind & Master      & Actor                   & 21K  & N/A                \\
P59 & 18-25 & Female         & Legally Blind & Master      & Student                 & N/A  & N/A                \\
P60 & 18-25 & Female & Legally Blind & Master      & Student                 & N/A  & N/A                                             \\
\hline
\end{tabular}
\end{table}



\subsection{Interview Design}

\rev{As noted earlier, this study is part of a larger project examining blind users’ broader perceptions and practices related to accessibility. As we continued conducting interviews, professional development repeatedly emerged as a central theme in participants’ TikTok use. In response, we added a dedicated section to the interview protocol that focused on this topic, which ultimately accounted for roughly one-third of the entire interview content.}

The professional development section of the interview protocol initially covered broad questions on job seeking and networking; it has been constantly revised based on participants' feedback throughout the study. The primary questions included: (1) \textbf{Overall}: e.g., what is your job and how do you use TikTok to help your career? (2) \textbf{Networking}: e.g., how do you build professional networks through TikTok? (3) \textbf{Monetizing}: e.g., how do you use TikTok to earn money? (4) \textbf{Advertising}: e.g., how do you advertise your business on TikTok? (5) \textbf{Getting Sponsorship}: e.g., how do you get sponsorship to make TikTok videos? (6) \textbf{Learning and Practicing}: e.g., what professional techniques have you learned or practiced through TikTok? (7) \textbf{Seeking jobs}: e.g., how do you try to look for a job through TikTok?

While we used these questions as primary prompts in the interviews, we asked follow-up questions to understand participants' situated experiences: (1) \textbf{TikTok affordances}: e.g., how do TikTok designs support or hinder your experiences? (2) \textbf{Accessibility}: e.g., what accessibility issues have you encountered on TikTok? (3) \textbf{Identity}: e.g., what role does your visual impairment play in the professional development process? (4) \textbf{Problem-solving}: e.g., how do you overcome the challenges, if any?

All the interviews were conducted remotely, via Zoom meetings or phone calls, based on participants' preferences. The interview sessions started with an introduction to the research team and the project. We proceed after getting the interviewees' consent. Then we verbally collected their demographic information and asked interview questions. The interviews ranged from 30 minutes to 90 minutes. We recorded the audio of interview sessions and transcribed the audio files into verbatim text transcripts. We removed all identifiable information, like name, address, and affiliation, to protect participants' privacy.

\subsection{Thematic Analysis}

To comprehensively capture participants' nuanced perceptions and practices and conclude the patterns and themes related to professional development on TikTok, the authors employed thematic analysis \cite{braun_reflecting_2019, braun_using_2006}. Data analysis was conducted by the first and third authors, with findings regularly discussed among the research team until consensus was achieved. The specific steps included the following: (1) \textbf{Immersive Reading}: We familiarized ourselves with the dataset by reading all the transcripts. We started reading all the transcripts line by line and got a basic understanding of each participant's experiences. (2) \textbf{First-round Coding}: We generated the first-round codes by open-coding all the transcripts, including all the data relevant to professional development. In this step, we paid attention to how participants demonstrated their specific professional development experiences and asked about their professional development goals and the role of TikTok. We coded as much as we could to capture a comprehensive picture of their professional development on TikTok. (3) \textbf{Code Grouping}: We grouped all the codes together to visualize their relationships to each other. After reviewing all existing codes, we compared their similarities and differences. For instance, some codes describe how participants use TikTok to show others their professional quality, and some show how they reach out to other TikTokers for job opportunities. For these codes, we grouped these codes as "networking." (4) \textbf{Code Refining}: We refined our groups of codes and generated themes based on the conceptual and thematic meanings of each group to ensure that, while the themes cover the overall experiences, each theme is mutually exclusive to the other. Especially, we refer to prior work on professional development on platforms \cite{ding_as_2022,corvite_social_2024} or by people with visual impairments \cite{saha_understanding_2020} when defining themes. \rev{For example, we got inspirations from prior work on general content creators' strategies and perceptions of professional development on social media platforms such as getting sponsorship and training skills \cite{ding_as_2022}; we also pay attention to accessibility issues that impact professional performance, such as productivity, as revealed by prior work \cite{saha_understanding_2020}.} (5) \textbf{Theme Refining}: We named all the themes and subthemes, making sure they represented the patterns of participants' experiences. \rev{We carefully compare all the code groups to make sure each theme and sub-theme mutually exclusive and present participants all aspects of professional development experiences. And the themes should be organized in a logical manner; therefore, we structure the themes as motivation, practices, and difficulties.} Finally, we have three themes, and each of them answers one of the research questions: the goals for professional development, the strategies for leveraging TikTok, and the challenges of using TikTok for professional development. (6) \textbf{Analysis Reporting}: Finally, we report the themes in Section \ref{results}. We provide an overview of themes and examples of codes in Table-\ref{tab:thematic-overview} and Table-\ref{tab:code}.

\begin{table}[htbp]
    \centering
    \caption{Overview of Themes}
    \renewcommand{\arraystretch}{2}
    \begin{tabularx}{\textwidth}{|
        p{\dimexpr 0.2\textwidth-2\tabcolsep} |
        p{\dimexpr 0.3\textwidth-2\tabcolsep} |
        p{\dimexpr 0.5\textwidth-2\tabcolsep} |
    }
        \hline
        \multicolumn{1}{|c|}{\textbf{Theme}} &
        \multicolumn{1}{c|}{\textbf{Subtheme}} &
        \multicolumn{1}{c|}{\textbf{Code}} \\
        \hline

        \textbf{Goals} &
        Overcoming Unemployment &
        Experiencing discrimination against blind people; Connecting TikTok to employment opportunities. \\
        
         &
        Promoting Career &
        Desire for career promotion; Connecting TikTok to potential promotion. \\
        \hline

        \textbf{Strategies} &
        Honing Professional Skills &
        Cultivating professional speaking skills; Citing algorithmic metrics for success. \\
        
         &
        Growing Professional Influence &
        Building own brands; Finding own niche; Expanding network. \\
        
         &
        Monetizing Content Creation &
        Receiving gifts from streaming; Getting payment from platform; Getting sponsorship from companies. \\
        \hline

        \textbf{Challenges} &
        Technical Challenges &
        Accessibility issues; Algorithmic bias. \\
        
         &
        Social Challenges &
        Online harassment; Unfair moderation. \\
        \hline
    \end{tabularx}
    \label{tab:thematic-overview}
\end{table}

\begin{table}[htbp]
    \centering
    \caption{Examples of Code}
    \renewcommand{\arraystretch}{2}
    \begin{tabularx}{\textwidth}{|p{\dimexpr 0.2\textwidth-2\tabcolsep} | p{\dimexpr 0.3\textwidth-2\tabcolsep}|p{\dimexpr 0.5\textwidth-2\tabcolsep}|}
        \hline
        \multicolumn{1}{|c|}{\textbf{Code}}
         & \multicolumn{1}{c|}{\textbf{\textbf{Definition}}} & \multicolumn{1}{c|}{\textbf{\textbf{Quote}}} \\
        \hline
        \textbf{Cultivating professional speaking skills} & BlindTokers aiming for using TikTok to train and increase their public speaking skills & "\textit{...And some of it was just me trying to get over my intimidation of speaking to people... because I was elected the president of our local [Blind Organization]. And I was using it [TikTok] to practice my speaking.} [P11]" \\

        \textbf{Finding own niche} & Blind content creators seeking specific area where they can better achieve professional success &  "\textit{I follow a lot of Disney content creators [who are sighted]... [I did videos to] compare rides as a sighted person versus a blind person. I think those are my favorites.} [P54]" \\
        
        \textbf{Algorithmic bias} & BlindTokers experiencing algorithmic oppression against blind users & "\textit{..when I’m trying to do dance, that’s when I get more shadowbanned. But I notice when I [just] use my voice, I get tons of views. So I’m like, okay, it doesn’t want to show people because I’m blind.} [P37]"  \\
        \hline
    \end{tabularx}
    \label{tab:code}
\end{table}

\subsection{Author Positionality}
The research team acknowledges that the authors' identities have shaped this study. The first author, a non-disabled man, was primarily involved in conducting the research, and the positionality statement primarily reflects his perspective. The other authors reviewed and supported this statement.

Grounded in social constructionism \cite{sparkes_narrative_2008}, I value human interpretation and acknowledge that my understanding of accessibility and visual impairment is shaped by academic training, literature, and—most crucially—the lived experiences shared by BlindTok participants. I recognize the privileges I hold, as academics, my access to information and resources, opportunities to voice my understanding, and the ability to reach broader audiences. I understand the privileges are not equally available to our participants. Such disparities prompt ongoing self-reflection: How does my identity shape my interpretations? And more importantly, how can this research authentically represent and benefit the blind community? To mitigate bias, I employ multiple strategies. First, I engage in reflexivity by embracing Tsing's concept of "contamination as collaboration" \cite{tsing_mushroom_2015}, a commitment to encountering and working across differences. Second, I follow Charlton's principle “nothing about us without us” \cite{charlton_nothing_2004}. This means ensuring the full and direct involvement of disabled individuals in all aspects of the research. Following Callus's recommendations \cite{callus_being_2019}, I prioritize listening to participants' narratives, conducting member-checking when possible, and consulting with accessibility experts and disability advocates to ensure that the study remains grounded in the realities of those it represents.

\section{Findings} \label{results}

\rev{As noted in the demographic section, aside from the students (10 out of 60), a sizable portion of participants were unemployed (13 out of 60). Although more than half of the sample was employed (37 out of 60), most reported incomes below the average per-capita income in the United States in 2023 (see Section Recruitment in Methods for details). These income patterns reflect broader employment challenges faced by many BlindTokers. Given the limitations of accurately assessing the income and employment status of people with visual impairments at a population level, it is difficult to determine whether BlindTokers are in a comparatively better professional position than visually impaired people at large. Nonetheless, all participants expressed a desire to use TikTok to improve their professional status and demonstrated clear forms of agency in their efforts to leverage the platform for career advancement. This also shows that the current study's primary focus is not to statistically compare BlindTokers' incomes with general visually impaired people, but to conceptually explore the patterns of BlindTokers' practices and desires for professional development with the socio-technical affordance of TikTok platform.}

\begin{figure}[htp]
    \centering
    \includegraphics[scale=0.5]{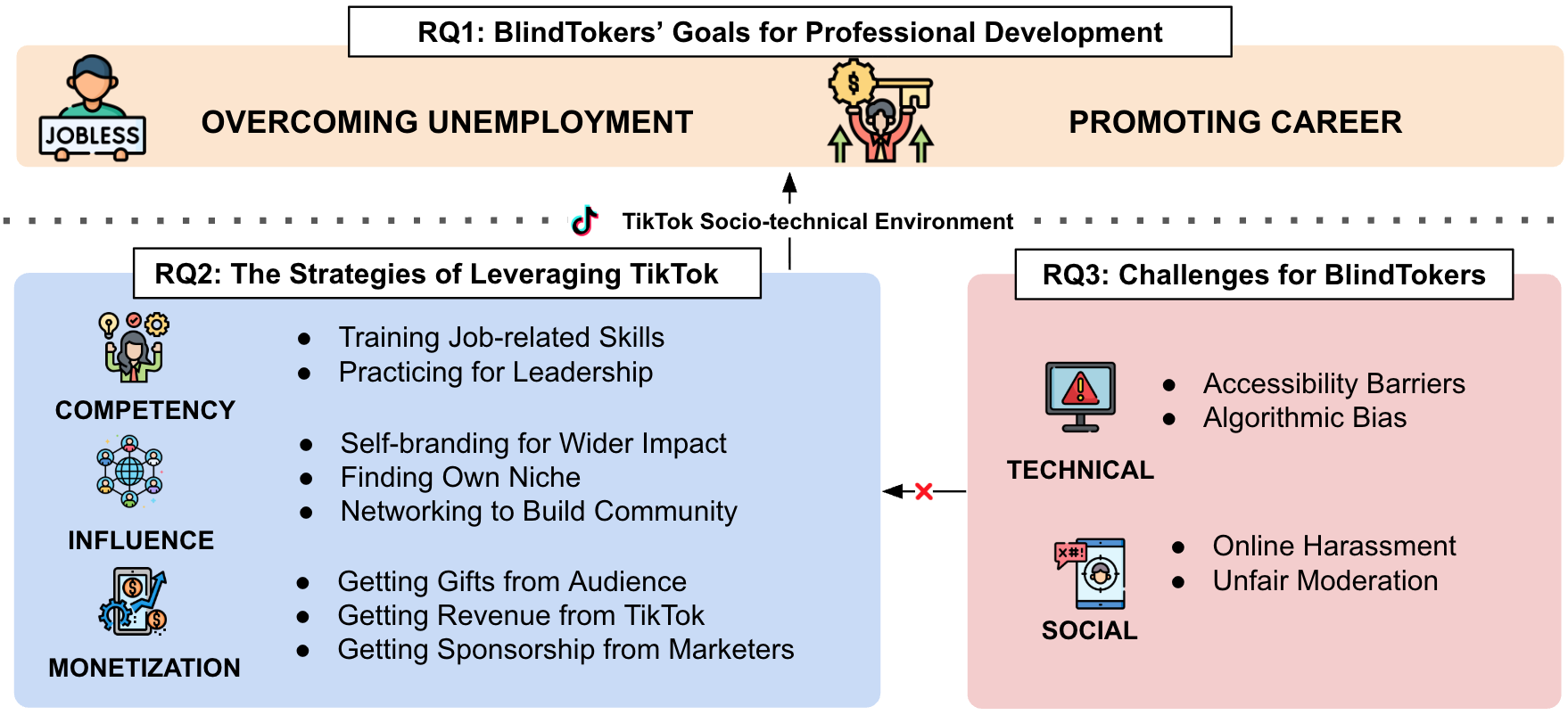}
    \Description{This is a Figure on BlindTokers’ motivations, strategies, and challenges in using TikTok for professional development. This figure consists of three thematic blocks—motivations, strategies, and challenges—that illustrate how BlindTokers leverage TikTok for professional development. Motivations block includes overcoming unemployment and promoting career. Strategies block, presented as supporting motivations, span competency building, network expansion, and monetization, all situated within TikTok’s socio-technical environment. Challenges block, including both technical and social challenges, are shown to hinder the strategies.}
    \caption{An Overview of the Findings: Goals of professional development include overcoming unemployment and promoting career. TikTok facilitates professional development with monetization (gifts, sponsorships, revenue), competency building (job skills, leadership), and networking (self-branding, community engagement). However, BlindTokers face challenges, including technical barriers, like accessibility issues and algorithmic oppression, alongside social challenges, such as online harassment and unfair moderation, during professional development on TikTok.}
    \label{fig:findings}
\end{figure}

Their professional development involved seeking employment, attracting customers, changing their professional tracks, or becoming professional TikTokers. Participants' practices included those that created career opportunities through learning, training, networking, and, in some cases, monetization. Our work uncovers how these experiences differ from their everyday interactions on TikTok, which are predominantly recreational and casual, as emphasized in prior work \cite{lyu_because_2024}. We organized the findings as three themes: BlindTokers' primary goals for professional development, their strategies for using TikTok for professional development, and the challenges they face (Figure-\ref{fig:findings}). The three subsections explore why BlindTokers engage in professional development, how they leverage TikTok to achieve their professional goals, and the obstacles they encounter. Their experiences reveal the complexities of navigating situations where disability identity, professional aspirations, and TikTok's socio-technical and algorithmic features intersect.




\subsection{BlindTokers' Primary Goals for Professional Development}

\rev{In the first subsection, we present BlindTokers’ goals for professional development. We identified two types of goals: fighting against unemployment and promoting career status. These goals reflect BlindTokers’ broader aspirations for their professional lives and can be understood as two stages within a continuous development process. However, our intention is to highlight the persistent vulnerability of blind people’s professional trajectories: whether employed or not, BlindTokers consistently expressed concerns about both job stability and opportunities for growth. We view them as coexisting challenges that many BlindTokers navigate simultaneously. In other words, a participant may strive to overcome unemployment while also aspiring toward career advancement.}

\subsubsection{Overcoming Unemployment}

As mentioned earlier, BlindTokers were experiencing unemployment; even some were employed, they were still making very low income. Many BlindTokers (20 out of 60) complained about the challenges of looking for jobs in real lives because of the discrimination against people with disabilities. For instance, P33 said: 

\begin{quote} \textit{...there’s a lot of us out there looking for jobs...It's tough to look for a job if you're blind because you're being discriminated against. And now I haven't had my fair share of interviews or I don't even get chosen for an interview... [even when I got one] I've sat through interviews where they’ve asked... "How would you get to work every day?"... It's frustrating because as a blind person you wanna progress, you want to be a part of society, you don't just want to be receiving the government's assistance your whole life. And I think it'd be great if we could find an opportunity through TikTok. [P33]} \end{quote}

P33's job search experiences involved extreme discrimination. In addition, her emphasis on "progress" and "being part of society" is consistent with previous research highlighting how people with disabilities do not want to be treated as an excluded part of society \cite{liu_i_2019}. P16 also shared his perceptions of TikTok and job seeking: 

\begin{quote} \textit{It's also a bit of a business thing because I'm a minister and I'm looking for a full-time church [job]. Now [when you apply for a job], you hand over your resume, and they want to check your Facebook, your TikTok, and everything you got. So if I can put out some quality material [related to Bible storytelling] that will make people smile, that will be kind of interesting. [P16]} \end{quote}

Similar to P33, P16 also considered TikTok as an important part of overcoming unemployment. He developed a theory about leveraging employers' use of social media in background checks: if the employer finds something that proves the competence of the applicant (storytelling as a church minister), it will increase their chances of being hired. Besides developing theories on leveraging TikTok for potential advantage, some participants actually benefited from the experiences related to TikTok when seeking jobs, P35 said: \textit{"I put TikTok on my CV... and especially because I do social media marketing and TikTok sort of falls into that quite nicely."}


\subsubsection{\rev{Promoting Career}}
In addition to employment, another major motivation for BlindTokers' professional development on TikTok is promoting their business. 37 participants had jobs or established their own businesses and desired for promotion or growth via TikTok. For example, P11 owned a gallery, and P13 was a financial consultant who earned around 1 million dollars before she retired. They also wanted to use TikTok as a platform to expand their impact. Take P11 as an example: 

\begin{quote} \rev{\textit{I'm pretty savvy. I do all of our social media for my business and I'm pretty good at it...I made one [video] that was a tour of our gallery...I thought I would post that on our Facebook page. And then I don't think Facebook really puts in their algorithm or whatever. [But] I have gotten a lot of comments from people on TikTok that they enjoyed seeing what a gallery [owned by a blind person] looks like... I'm waiting for the day when somebody just recognizes me on the street and says, I've seen you on TikTok! [P11]}} \end{quote}

\rev{P11 was aware of the importance of using social media to advertise their local business. Her comparison of Facebook and TikTok indicated that TikTok was much more successful in terms of reaching out to a larger audience.} In addition to advertising local businesses, some participants also worked on managing social media content on multiple platforms, P16 shared:

\begin{quote} \textit{And that's another message of the [YouTube Channel name], like getting visually impaired, it's not like a death sentence...You can still do these things, and we're gonna show you that you can...because a lot of people don't know that our YouTube channel exists. I figured out TikTok to be a good avenue just to do that. But at the same time, get people to find the YouTube channel. [P16]} \end{quote}

P16, a radio host, had a YouTube channel dedicated to sharing inspiring and instructional information for audiences who are also visually impaired. Like P11, P16 also realized the success of TikTok in terms of reaching out to more people, especially people with visual impairments. P16's experience also emphasized the importance of creating sources to help the community that experiences the loss of instruction or motivation to become successful in their lives. Reaching out to and benefiting people with visual impairments was one of the most important reasons for growing businesses on TikTok.

\subsection{Strategies of Leveraging TikTok for Professional Development}

As briefly introduced in the last subsection, many BlindTokers believed TikTok could help their professional development by finding jobs or attracting customers. In this section, we dive into the strategies of leveraging TikTok for professional development, elaborating on BlindTokers' specific strategies of professional development using TikTok, including monetizing for profits, honing skills for competency, and socializing for networking purposes. We also illustrate the specific socio-technical features, designs, and functions of TikTok that were used by BlindTokers for professional development. 

\subsubsection{Honing Professional Skills}

One of the major subthemes of utilizing TikTok for professional development is honing professional skills, as reported by about half (31 out of 60) of the participants. Skill development is one of the most important parts of professional training, as it builds up one's core competency and can significantly increase one's chance of professional success \cite{cherubini_elucidating_2021}. Moreover, BlindTokers' skill development has a lot to do with leveraging or navigating TikTok's socio-technical affordances. A popular example is practicing speech skills on TikTok. According to P11:

\begin{quote} \textit{And some of it was just me trying to get over my intimidation of speaking to people... because I was elected the president of our local [Blind Organization]. And I was using it [TikTok] to practice my speaking. [P11] }\end{quote}

Many participants told us that, making speeches is an important skill in many professions, like being a teacher, church minister, business owner, or organization leader. Good speaking skills, like storytelling and idea conveying, could significantly enhance one's communication and leadership. This could also lead to promotion to managerial positions \cite{dancisinova_presentation_2017}. 

In addition to using TikTok for training professional skills, some participants also cited the algorithmic metrics when evaluating their training performance. For instance, P16, who was looking to become a church minister, continued to post videos about Bible stories on TikTok:

\begin{quote} \textit{\rev{Now [applying for a job], you hand over your resume, they want to check your Facebook, TikTok, everything you got. So if I can put out some quality material that will make people smile, that will be kind of interesting...And part of the idea to TikTok is connecting people. Now my feed is full of blind people, and other progressive Christian ministers. And that's been kinda fun to see. Because my brother is a school principal, earns about \$80,000 per year, has a car, has a house. But when the kids see me, “Wow, uncle, you have 1500 followers on TikTok!” That's what's cool for the kids these days. So you got to keep your reputation on them. [P16]}}\end{quote}

\rev{P16 believed that creating Bible-story-oriented content on TikTok and other social media platforms could improve his chances of securing a full-time ministry position, because social media influence (evaluated by algorithmic metrics) signals professional competence as a storyteller. He treated his number of followers as an indicator of his storytelling success on TikTok. His enjoyment when hearing his nephews praise his TikTok popularity also suggested that, in this new era, professionalization on TikTok and in traditional contexts (for example, his brother being a high school principal) were equally meaningful markers of success.}


\subsubsection{\rev{Growing Professional Influence}}

Another strategy for leveraging TikTok for professionalization is growing professional influence. The first part of growing professional influence is self-branding. Specifically, BlindTokers introduced how they used TikTok for self-branding, utilizing the popular styles of video making on TikTok so they could increase their personal impact on the audience. One strategy was to keep posting videos with personalized features. P13, a former financial consultant who asked her sighted friends to operate her own TikTok account, shared how she built her personal brand:

\begin{quote} \textit{You can see the writing on my videos and my cover things are pink. That's all me. And so that the people [sighted friends]... they know I like bling and emojis. And so definitely my flare...So I just wanted to create my brand. So I say that every time, just to give people the impression. [P13]}\end{quote}

P13's experience indicated that she saw the importance of keeping her own personality when building TikTok accounts for networking; she was also experienced in asking people in her network to help her make content to meet her personality standards. P13's request for assistance from her friends also suggests accessibility barriers on the platform.

\rev{In addition, some participants, when reflecting on the importance of self branding and the diverse background of content creators on TikTok, decided to find their own niche based on their experiences as people with visual impairments.} For instance, P54, an actress, used TikTok to share her insights on the entertainment industry:

\begin{quote} \textit{I follow a lot of Disney content creators. That's originally why I started watching TikTok, [some videos by sighted creators]...so she's almost right. But that's not the exact truth [because the videos did not cover perspectives from people with visual impairments]... [I did videos to] compare rides as a sighted person versus a blind person. I think those are my favorites. [P54] }\end{quote}

P54 surveyed videos on TikTok about Disneyland experiences. She found that most videos only talked about people without visual impairments. She wanted to bring people with visual impairments' perspectives into videos that evaluate their Disneyland experiences. This became her unique niche of content creation. 

Lastly, some participants shared how they networked on TikTok and benefited from it. P58, a former actor, also worked in the entertainment industry. He lost his vision in the middle of his career and described his career transition:

\begin{quote} \textit{I have gone from an enabled actor to a disabled actor. What steps should I take to be sure that I can continue my career? I actually follow a lot of casting directors... there's one... And she says, you might want to do this (voiceover job)...[then] I follow quite a few voice actor people...and I keep getting told that I'd be really good with voice. [P58]}\end{quote}

P58 experienced an unexpected professional change. He used TikTok to get advice from more experienced people to help his professional transition. His experience also suggested that his voice-acting competency contributed significantly to his successful transition.

In addition to networking for advice, some participants also shared how they provide advice or share expertise with people on TikTok. P44 was a teacher who liked to post TikTok videos to share with others:

\begin{quote}\textit{I think my favorite is just my videos with my assistive technology and just showing them how I use the assistive technology in the classroom, because people don't realize just how advanced technology is...So I get a lot of teachers messaging me and asking more about different types of technology because it could help another student in their class. [P44]} \end{quote}

Teaching with assistive technology benefits a wide range of people, including teachers and students with visual impairments. P44 actively shares this information with people on TikTok to help other teachers perform better in such professional contexts and to give students with visual impairments a better education.

\subsubsection{Monetizing Content Creation}
Lastly, while many BlindTokers leverage TikTok’s traffic to support business opportunities beyond the platform, some of them (15 out of 60), especially professional TikTokers, generate income directly through TikTok content creation. Monetization through content creation is a necessary part of professional development on platforms \cite{choi__2022}. Participants shared how they learned from the current monetizing methods from algorithmic platforms and used them on TikTok to earn revenue. One way is to receive gifts from live streaming. P11 shared a story about her friend, who was also a BlindToker:

\begin{quote} \textit{He likes to sing. He's all that talented. But he's always talking about his life and he sings with a karaoke machine. And I think he said he got close to \$40 last night. So while you're live streaming, people send you like a little diamond or gifts. And it turns into money somehow for your account, he's got a huge number of followers. [P11]} \end{quote}

P11's friend's success indicated the competency of content creation. The large number of followers that led to monetized revenues also suggested the importance of using algorithmic metrics. Another method of monetization was to get paid by the platform. Many platforms provide incentive programs that reward content creators' revenue based on the quality of their content. P38 told us:

 \begin{quote} \textit{I'll see how it goes and ever can reach 10,000 subscribers in a year. I hit just over 8,000 in a year, which was fantastic. But it's a lot of work. It really is a full-time job. It really is if you do it properly and you’re capitalizing and making good content. It takes up about a 3-day process from recording to editing a video about to be ready for release. It's so time-consuming... [P38]} \end{quote}

P38 was a professional content creator, and his work quality was directly associated with his income from TikTok. His emphasis on the number of subscribers again showed the importance of algorithmic metrics when evaluating the success of a professional content creator. 

\rev{In addition to the audience and the platform, some participants mentioned getting revenue from sponsorship. P58, the aforementioned actor who used TikTok to get advice to become a voice actor, told us he got sponsorship to make TikTok videos: \textit{"I did some sponsored videos...One was for a company and disability wheels. So they were already quite aware of access requirements and things."} In P58's experience, his professional growth was accompanied by getting sponsorships from companies specializing in accessible equipment for people with disabilities, suggesting a suitable and sustainable monetization source for BlindTokers.} 

\rev{The compliment was echoed by all participants, as they all appreciated sponsors' considerations for accessibility during collaboration. That said, some of them complained about the control over content from some companies. For instance, P31 complained when she received strict requirement from sponsors:} 
\rev{\begin{quote}\textit{"Sponsors don't always understand TikTok as a platform and have an idea of how it's supposed to work...then we have to follow the script, to use this caption, these hash tags, then our audience would be like, this doesn't sound like the you" [P31]}
\end{quote}}
\rev{This suggests a more nuanced situation where some BlindTokers have limited control over their creative content when working with sponsors, which may undermine their long-term professional development as content creators. Relying solely on sponsor-provided scripts can constrain their authentic voice and potentially lead to negative perceptions from their audience.}

\subsection{Challenges for Professional Development on TikTok}
Although BlindTokers were motivated about professional development and experienced in employing specific strategies, they encountered numerous challenges on the platform. Overall, we identified two categories of challenges that significantly hindered BlindTokers' professional development experiences: technical and social challenges.

\subsubsection{Technical Challenges}
Technical challenges on TikTok arise from its design (particularly in terms of accessibility) and algorithmic management. According to almost all the participants (55 out of 60), these challenges not only hindered user experience but also negatively impacted their professional development. Many BlindTokers complained about the technical issues in content creation, as it affected the quality of content they could create. For instance, P15 said:

\begin{quote}\textit{...when you have all the filters at the bottom, I don't understand what any of those little icons, the little images mean. It's more of a contrast, a font size, and no accessibility as far as I can tell. Those are the main issues. [P15]}\end{quote}

Accessibility issues on social media have been reported by numerous studies, as they harm users' experiences \cite{lyu_i_2024,lyu_because_2024}. In the face of accessibility issues, many participants were concerned about making mistakes in content creation, which could further affect their professional performance. For example, P14 expressed his concerns when planning to have a TikTok account for his business:   

\begin{quote} \textit{I didn't want to start posting as a business page on TikTok unless I really have the platform figured out 100\%, because the business is far different from a personal account...(there is) less margin of error as a business. [P14]}\end{quote}

While P14 recognized the benefits of creating TikTok content for his business, accessibility issues that could impact content creation ultimately hindered his plans. In this situation, some BlindTokers who wanted to create high-quality work had to conduct extra work to overcome the accessibility issues. For example, P17, a music composer, shared how he made music content on TikTok:

\begin{quote}\textit{(After finished recording, I will save it to my computer's file...I use a keyboard because the interfaces are not accessible to people who use VoiceOver. Pretty much from clipping it down to less than a minute, setting the picture of the video, matching the length with the audio file, sending it to my phone, and on my phone I will hit the aspect ratio button. It's something like horizontal and vertical and then hit 9:16. So it's like standing vertically. [P17]}\end{quote}

P17 described the meticulous process of arranging various resources, managing multiple devices, and navigating different standards to create high-quality music. His efforts underscored the accessibility barriers that TikTok imposed on BlindTokers.

In addition to accessibility barriers, BlindTokers also complained about the algorithmic oppression on TikTok against creators with disabilities. For instance, P37 said:

\begin{quote}\textit{I've been thinking... reaching out to these companies and I could just talk about their product on my TikTok. I would love to be sponsored ...when I'm trying to do dance, that's when I get more shadowbanned. But I notice when I [just] use my voice, I get tons of views. So I'm like, okay, it doesn't want to show people because I'm blind. [P37]} \end{quote}

This is consistent with prior work on how algorithmic platforms limit the visibility of content created by users from minority groups \cite{lyu_i_2024, simpson_how_2022,boffone_jewishtiktok_2022}. In this case, algorithmic bias could also significantly hinder the chance of getting more views and therefore, decrease the chance of successful networking or monetization.

\subsubsection{Social Challenges}
In addition to technical challenges, we also observed from about half of participants' (28 out of 60) stories that they encountered social challenges. Unlike challenges that are driven by technical issues, social challenges refer to those induced by issues like discrimination and fairness issues. The most significant issue was online harassment. P44, the aforementioned teacher who enjoyed networking with people on TikTok and sharing expertise in terms of using assistive technology in class, told us a story:

\begin{quote}\textit{I had a really bad experience with this little boy...he just kept commenting "You're not blind." I'm just like "Why?"... people say it all the time, and it's very frustrating. Once, I made a video about how I used TikTok. I was showing how I use VoiceOver and stuff, and this person literally commented, "But if you're blind, how are you using TikTok?" I was like, "I'm literally showing you in the video!" [P44]} \end{quote}

As P44 had mentioned, she used TikTok to exchange professional skills (teaching with assistive technology in class) with her colleagues. However, the online harassment, stemming from the ignorance of people with visual impairments' lifestyles, significantly frustrated her experiences in exchanging professional skills on TikTok. While BlindTokers wanted to block the trolls, the accessibility issues would hinder the blocking experience, as reported previously \cite{lyu_i_2024}. 

In addition to online harassment, other BlindTokers further reported how they were banned due to the unfair treatment of people with visual impairments. Many participants shared that they encounter unfair moderation practices due to systemic discrimination against people with visual impairments. For instance, P42 a content creator who focused on raising awareness of creating an accessible environment for people with visual impairments, shared how her content was moderated by TikTok:

\begin{quote}\textit{...[in one of my videos] the comment was, "What circus did you crawl out from?" And I made a response video, and the description of the video, it said the circus where I f**ked your mom...I don't actually talk like that, but like everybody needs to know like nobody could take your power from you...He ended up reporting my comment back. And that video ended up getting taken down. I was pissed because it was one of my highest-ranked videos. [P42]} \end{quote}

P42 shared an experience in which she was offended by someone using ableist language. However, when she defended herself, she was the one who faced punishment. This not only led to her video being moderated but also negatively impacted her algorithmic metrics, as a high-ranking video was taken down. Unlike P44, whose frustration stemmed from interactions with trolls, P42’s experience highlighted TikTok’s lack of attention to harassment against people with disabilities.

\section{Discussion}

Our findings present a comprehensive picture of professional development among people with visual impairments: (1) Participants expressed a strong desire for professional growth, motivated by both personal significance (“as a blind person you wanna progress, you want to be a part of society” [P33]) and economic needs. Yet they faced systemic barriers in traditional professional development contexts, including limited job opportunities, discrimination, and workplace accessibility challenges. As a result, they turned to alternative approaches (e.g., TikTok) to pursue opportunities. (2) On TikTok, participants identified strategies for advancing their goals, including training (to improve job prospects), networking (to explore both conventional and emerging opportunities), and monetization (to enter new professions and diversify income sources). However, they continued to encounter numerous sociotechnical challenges shaped by TikTok’s environment, requiring ongoing effort to overcome these obstacles.

\rev{We also identified two types of professions among BlindTokers: while some used TikTok to support their traditional careers, such as business owner, church minister, or teacher, others developed new professions as content creators. These two categories are not mutually exclusive. Some traditional professionals relied on TikTok content creation to reach wider audiences and expand their networks, while some content creators simultaneously sought opportunities in more traditional career paths. This illustrates the flexibility that TikTok provides for BlindTokers’ professional development.}

Inspired by BlindTokers’ diverse understandings and practices of professional development, we revisit the concept in our discussion and propose an alternative approach. Our intention is not to present TikTok as superior to traditional professional development pathways, but rather to highlight how BlindTokers leverage the platform to pursue professional goals when conventional contexts fall short (thus shaping an “alternative” form of professional development). We further examine the skills involved in this alternative approach and outline design implications to better support it.

\rev{To be noted, this paper does not claim that TikTok replaces traditional education or disability-specific training. For instance, we found that TikTok use did help BlindTokers diversify their revenue sources, but we did not find that TikTok necessarily increased their overall income than traditional contexts. Rather, aligning with disability studies and HCI research showing that people with disabilities benefit from receiving support across multiple infrastructures, including traditional ones (e.g., schools) and nontraditional ones (e.g., social media, online forums) \cite{atwell_left_2025,arnold_beyond_2025}—we view TikTok as a complementary, informal learning site that operates alongside formal channels. It supports BlindTokers in areas where traditional systems fall short in accessibility, representation, or continuity.}

\rev{That said, we do not view TikTok as a perfect solution, even as a complementary one. We also highlight numerous social and technical limitations of the platform. It is precisely these gaps that draw attention to the emerging professional skills (Section 5.2) BlindTokers have developed to make professional development on TikTok possible, smoother, and more beneficial. Again, our aim is to present TikTok as a meaningful new perspective, not as a replacement for or superior alternative to traditional approaches.}

\subsection{Revisiting Professional Development in the BlindTok Case: An Alternative Approach}

\rev{We revisit the concept of professional development, arguing that the traditional definitions of it are problematic and discriminatory. As stated in the related work section, traditionally, institutions hold power and privilege in defining professional training, skills, and qualifications \cite{greenwood_attributes_1957}. The institutionalized conception of professional development is problematic, as the monopolized power in professional development leads to the marginalization of people with disabilities. As reported by prior work \cite{saha_understanding_2020} and BlindTokers in the current study, people with visual impairments encountered tremendous professional development challenges in the current society. People with disabilities face the educational and professional standards set by institutions with little consideration of accessibility. Specifically in our study, BlindTokers experienced inaccessible training and education, and discriminatory job interviews.} 

Recently, the rise of social media platforms, like TikTok, has provided an alternative approach to professional development. Compared to traditional approaches, this alternative one allows BlindTokers to freely adjust their professional paths, to inclusively incorporate their disability identities into professional practices, and to flexibly work for revenue. Recent HCI research has paid attention to the professional development on social media platforms \cite{ding_as_2022, ma_multi-platform_2023, corvite_social_2022}. Researchers specifically investigate how platforms' socio-technical affordances enable individualized professional development. This line of work on professional development on platforms has emphasized the importance of community-engaged and interests-driven approaches among platform users (which is an important "distinction from highly institutionalized \cite{ding_as_2022}" approaches). With the platforms, users can endow meanings, norms, and significance based on their individualized situations, interests, and identities when working for professional growth. 

Following this line of work, our findings expand the literature by adding a case of people with visual impairments on TikTok and presenting their alternative approach to professional development. Therefore, we propose the notion of \textbf{alternative professional development}
, a type of approach to professional growth that is complimentary to traditional and institutional pathways, where individuals use socio-technical environments (especially emerging platforms like TikTok) to direct their careers, redefine competency, adopt personalized measures of success, and diversify revenue streams. This approach is formed with both BlindTokers' agency and the TikTok platform's socio-technical environment. We unpack each dimension of the notion of alternative professional development as follows:

\begin{enumerate}
    \item \textbf{Greater Control over Career Paths.} As reported by participants, they could use TikTok to promote their current professional status (e.g., attracting customers for current businesses), to change their professional tracks (e.g., transition to voice actors after being visually impaired), or to try emerging professions (e.g., becoming professional TikTokers). The flexibility allows participants to circumvent the constraints set by the traditional and institutionalized employment routines. 
    \item \textbf{Redefines Disability as Competency.} In institutionalized professional settings, disabilities are often framed as deficits and defects, which wrongly suggests a lack of competency \cite{castro_experiences_2024}. In our study, BlindTokers, through their endeavors, translate their visual impairments into professional competency. The endeavors include identifying niches (e.g., content creation from a person with visual impairment's perspective), discovering talents (e.g., becoming voice actors or singers), and exchanging expertise (e.g., sharing assistive technology knowledge). The finding echoes earlier work on leveraging disability identity in platform-based sales promotion \cite{borgos-rodriguez_understanding_2023}. Such endeavors further helped BlindTokers tremendously in building professional networks and securing professional positions. \rev{We acknowledge that this strategy can carry stigmatizing effects against BlindTokers, as any form of identity leverage can. However, in content-creation contexts, recognizing and drawing on one’s identity is a widely used strategy for building an audience and forming networks around shared interests, especially among marginalized populations \cite{simpson_rethinking_2023,simpson_hey_2023,lyu_because_2024}. Prior research \cite{lyu_i_2024} also shows that community building and continued voicing out (e.g., creating content tied to one’s identity) can help mitigate bias and address misconceptions held by people outside the community or identity group.}
    \item \textbf{Individualized Pathways}. Traditionally, professionals are expected to serve specific communities through their expertise and skills, and their services should be deemed valuable within particular contexts \cite{barber_problems_1963}. However, adopting a bottom-up approach, BlindTokers articulate their sense of professional success from a personal perspective. As some participants shared, their professional efforts are often directed toward supporting peers who face similar professional or life challenges. In addition, some participants also evaluate their professional success based on algorithmic metrics instead of conventional metrics like salary.
    \item \textbf{Diversification of Revenue.} Participants reported a wide range of income sources, including TikTok's incentive programs, audience gifts, and sponsorship commissions. This diversity allows them to strategically find their own niche and allocate professional efforts toward the most suitable and sustainable revenue streams. In this context, their identity as visually impaired creators plays a crucial role; it not only shapes their content but also helps foster a community that connects them to potential paying audiences and sponsors.
\end{enumerate}

\rev{Reflecting on BlindTokers’ alternative approaches to developing professional status, we advocate for a more inclusive conceptualization of a profession, rather than adhering to traditional definitions such as “a vocation founded upon an understanding of the theoretical structure of some department of learning or science” \cite[p.~107]{cogan_problem_1955}. The inclusive conceptualization of profession should embrace accessibility, flexibility, and adaptation, enabling people to explore professional pathways in environments that support their abilities and strengths rather than judge their limitations. As a result, we further argue for an \textbf{alternative professional development} that, in the context of BlindTokers, recognizes and validates platform-enabled practices, including content creation, experience sharing, and community building, as legitimate professional endeavors.}

\subsection{Examining the Alternative Approach to Professional Development: Emerging Professional Skills}

In the previous section, we introduced an alternative approach to professional development based on BlindTokers' cases and advocated for a more inclusive definition of professional development. In this section, we continue to discuss the alternative approach by identifying emerging professional qualities. Proper recognition of work, skills, and labor is essential in professional settings, whether within traditional or alternative pathways. Prior work has emphasized the importance of applying equitable models when evaluating people with disabilities' professional performance \cite{kafer_feminist_2013}. For people with disabilities, failing to do so leads to perpetuating ableism in workplaces, resulting in social justice problems. 

As presented in the findings section, BlindTokers' professional development was driven by their situated desires, perceptions, and practices; the professional development was also mediated and shaped by the TikTok platform. More importantly, while taking the alternative approach, they still face a number of challenges and have to navigate the platform barriers, such as inaccessible interface designs, algorithmic biases, and online harassment. To overcome these obstacles and ensure their professional performance, BlindTokers had to conduct tremendous additional work. And the work was mostly invisible and unrecognized when BlindTokers' professional performance was evaluated. The lack of recognition leads to an examination of professional qualities in the alternative approach to professional development on TikTok.

Prior research has examined the types of labor performed by content creators on digital platforms, including creative labor \cite{simpson_rethinking_2023} and algorithmic labor \cite{ma_how_2021}. While the former highlights creators’ engagement with audience communities, who serve as important sources of learning, networking, and monetization, the latter focuses on the efforts required to understand and navigate algorithmic evaluations of professional performance. However, only a few studies have explored these forms of labor from the perspective of people with visual impairments. One study of a visually impaired community on video-sharing platforms \cite{lyu_i_2024} foregrounds both accessibility and emotional labor, emphasizing the importance of overcoming accessibility barriers and sustaining a positive emotional presence during content creation. Yet, this study does not link these forms of labor to users' professional development. Building on prior work and the findings of the current study, we examine the inherent standards of the alternative professional development approach on TikTok and highlight the qualities that are particularly valued within this context:

\begin{enumerate}
    \item \textbf{\rev{Accessibility Work}} \cite{lyu_i_2024,branham_invisible_2015}. The most fundamental standard is accessibility work. BlindTokers who sought to search for jobs, network for opportunities, or create monetized content first had to overcome significant accessibility barriers. As reported in our study and prior work \cite{lyu_because_2024,lyu_i_2024,lyu_i_2024-1}, these barriers include inaccessible content (e.g., images or videos without proper audio descriptions), interfaces (e.g., icons, buttons, and stickers incompatible with screen readers), and interactions (e.g., overwhelming floating comments in live streams). Notably, accessibility work is required in all of the subsequent categories.  
    \item \textbf{Algorithmic Literacy} \cite{ma_how_2021}. As repeatedly mentioned by participants, algorithmic metrics, like the number of followers and likes, were the key to professional success for content creators. However, the algorithmic evaluation was oftentimes affected by not only the accessibility barriers but also the external factors like algorithmic biases or moderation, and online harassment. The challenging situation required BlindTokers to have high algorithmic literacy when pursuing professional success. 
    \item \textbf{Advocacy Skills} \cite{sannon_disability_2023,kaur_challenges_2024,chen_voice_2025}. As members of a marginalized population, BlindTokers emphasized the role of advocacy in their professional development. For example, some intentionally created videos from the perspective of people with visual impairments or built personal brands based on their identities to address the lack of representation on TikTok.
    \item \textbf{Emotional Labor} \cite{lu_emotional_2022, dosono_moderation_2019}. A few cases mentioned that BlindTokers had to maintain a stable and calm mood when confronted by trolls. For example, P44 shared how she kept professional and patient when someone harassed her under a video she showed how to use assistive technology to her colleagues.   
\end {enumerate}

The alternative approach to professional development, as presented by BlindTokers, draws attention to a new skill framework. This framework demonstrates not only how to acquire digital skills but also how to develop capacity and adaptation. The framework further points out key criteria for professional evaluation in the emerging context, where people with visual impairments conduct professional development on TikTok. To better translate our discussion into practical outcomes, we also offer specific design implications to support the alternative approach.

\subsection{Design Implications: Supporting the Alternative Approach to Professional Development}

We must build an inclusive and equitable platform to support BlindTokers' alternative professional development. This involves a revolution in the work of designers, platform managers, and accessibility advocates. The core mentality of design implications is to address socio-technical barriers that hinder BlindTokers’ professional growth. Practical approaches should work on facilitating skill-building, fostering networking, and supporting monetization. We provide several specific implications as follows. 

\begin{enumerate}
    \item \textbf{Professional stakeholders on TikTok should recognize BlindTokers' accessibility work as professional expertise.} BlindTokers conduct professional development with various stakeholders, including the TikTok platform, peers, audience, and sponsor companies. To overcome accessibility issues, BlindTokers have to exert extra labor. When evaluating BlindTokers' professional performance, these stakeholders should take accessibility into consideration, valuing it as a professional skill instead of an individual burden. In addition, TikTok designers can also foreground such labor by highlighting content creators' accessibility contributions with algorithmic ranks to increase the visibility of accessibility contributors. The TikTok platform can also host events like platform campaigns to foreground such labor to call for attention to the recognition of BlindTokers' work in professional growth.

    \item \textbf{Sponsors should develop and apply equitable monetization models for BlindTokers}. Getting monetized rewards is an important part of professional development. BlindTokers, when creating content, encounter numerous challenges, like accessibility barriers and algorithmic bias. These challenges affect their interactions with the audience, content quality, reception of rewards, and visibility to potential sponsors. To overcome the challenges, there are multiple approaches. First, the TikTok platform should ensure that monetization tools, such as gifting and revenue management, are accessible. Second, when evaluating BlindTokers' professional performance, like content quality, monetization sources like the platform, audience, and sponsors should apply tailored models for BlindTokers, e.g., considering the aforementioned accessibility labor, and avoid using a one-size-fits-all model. These could ensure that BlindTokers have equal access to economic and professional growth.
    
    \item \textbf{The TikTok platform should build an inclusive and safe environment for professional networking}. Networking is also an indispensable part of professional development. However, BlindTokers reported that during their networking, they faced various problems like moderation, harassment, trolls, and ableism. These problems isolated BlindTokers and frustrated their passion for networking with others. The TikTok platform should revise its content moderation policy, including providing more contextual moderation mechanisms and more accessible, appealing approaches for moderation cases that involve BlindTokers. The TikTok platform should also remove ableist speech. TikTok can also incorporate designs like group chats, where BlindTokers can safely communicate, learn, and network with their professional collaborators.

    \item \textbf{Public sectors and disability organizations should provide accessible training resources on TikTok}. Recognizing that there is a growing number of BlindTokers who use TikTok for professional development, public sectors, and disability organizations should actively prepare and disseminate accessible training materials (such as tutorials on using screen-readers to create content) on TikTok. This could utilize the algorithmic power of TikTok, such as hashtags, to reach out to the target audience more effectively. These organizations can also conduct research programs to understand BlindTokers' professional desires in the TikTok context and, therefore, develop more suitable and up-to-date materials for professional training and learning. The efforts by public sectors and disability organizations can also help TikTok build a diversified creator base, increase user retention, and enhance public image.

\end{enumerate}

\section{Limitations and Future Work}
While our study offers valuable insights into the professional development of BlindTokers, we acknowledge several limitations. First, our participant sample was exclusively drawn from active TikTokers, which provided rich, platform-specific insights but may limit the generalizability of our findings to other social media platforms (or users that do not actively share their experiences). Given the diversity of socio-technical designs across platforms like X, Instagram, and YouTube, each with distinct algorithmic systems, content moderation policies, and community structures, users with visual impairments may encounter different challenges and opportunities elsewhere. Future research could explore the professional development of blind users on other platforms or examine cross-platform dynamics to gain a broader understanding of how different environments shape their professional experiences.

Second, our data was collected during a specific time period (Summer 2022 to Spring 2023). Given the dynamic nature of social media platforms, TikTok's features, such as accessibility barriers, algorithmic biases, audience engagement mechanisms, and monetization policies, are subject to frequent updates. As a result, some of our findings, discussions, and design implications that reflect platform-specific characteristics may not fully apply to newer versions of TikTok. We encourage future researchers to conduct longitudinal or comparative studies that track changes over time, in order to better understand how evolving platform designs shape users' professional development experiences.

\rev{Third, this paper’s focus is limited to exploring the perceptions and practices of BlindTokers’ professional development in a new context (TikTok). And the data analysis is based on what participants have shared with us. Therefore, many types of data, such as the specific association between income and occupation of BlindTokers, and more types of goals of professional development (e.g., career advancement, salary growth) may not be mentioned in the participants’ reports. Future researchers joining this line of work can further work on such research directions and produce more meaningful insights.}
\section{Conclusion}
We conducted an interview study with 60 TikTokers with visual impairments (referred to as "BlindTokers") to explore their experiences of professional development on the TikTok platform. The study identified three major themes: BlindTokers' motivations for pursuing professional development, the strategies they employed to leverage TikTok for professional growth, and the challenges they faced along the way. These findings revealed an alternative approach to professional development that contrasts with traditional models. Based on our results, we advocate for a more inclusive conceptualization of professional development and offer design implications to better support this alternative path.

\begin{acks}
Thanks for the reviewers' comments. 
\end{acks}

\bibliographystyle{ACM-Reference-Format}
\bibliography{Reference/Yao}
\end{document}